\documentclass[a4paper, amsfonts, amssymb, amsmath]{article}
\usepackage[utf8]{inputenc}
\usepackage{color}
\usepackage[dvipsnames]{xcolor}
\usepackage{xfrac}
\usepackage{cite}

\usepackage{appendix}
\usepackage{soul}
\usepackage{ulem}
\usepackage{graphicx}
\usepackage{soul}
\usepackage{authblk}
\usepackage{amsmath,amssymb,amsfonts,graphicx}
\usepackage[a4paper,top=2cm,bottom=2cm,left=2cm,right=2cm]{geometry}
\usepackage{tikz}
\usepackage[pdftex, pdftitle={Article}, pdfauthor={Author}]{hyperref}
\usepackage{lscape}

\begin{document}

 \vspace{1.5cm}
 
\begin{center}
{\bf\Large Under the spell of chiral symmetry}

\vspace{0.8cm}

Vicente Vento \vskip 0.7cm

{\it  Departamento de F\'{\i}sica Te\'orica and Instituto de
F\'{\i}sica Corpuscular}

{\it Universidad de Valencia - Consejo Superior de Investigaciones
Cient\'{\i}ficas}

{\it 46100 Burjassot (Val\`encia), Spain, }

{\small Email: vicente.vento@uv.es}

\end{center}
\vskip 1cm \centerline{\bf Abstract}
In this reflection, I want to look back over more than thirty years, from 1978 to 2008, of my professional and personal relationship with Mannque Rho. I will briefly describe some of the original ideas that emerged from our collaboration: the Little Bag, the Cheshire Cat principle, and Skyrmion matter. I do not intend to be exhaustive, neither in content nor in references, which can be found in our published works. Instead, I want to retrace the path we took together through physics, accompanied by many good friends.

\vspace{1cm}


\vspace{1cm}

\section{Introduction}
\label{sec1}

I met Mannque Rho in 1978 during his sabbatical visit to Stony Brook. At that time, I was a graduate student in the group of  Gerry Brown. Gerry and Mannque had been discussing that the MIT bag model explicitly violated chiral symmetry in its boundary conditions. They had written a paper full of very valid arguments and reasonings, "The Little Bag," but which lacked a consistent technical foundation \cite{Brown:1979ui}. Together, the three of us formalized these ideas in a work we titled "Little Bag Dynamics" \cite{Brown:1979ij}.  During the following two years, I continued working on the formalism and the consequences of the ideas from that first paper, and Mannque, already from Paris, remained in contact and interested in my developments \cite{Vento:1980mu}. It was precisely in discussions with Mannque that we realized the non-conservation of baryon number in my solitonic solution, which some years later gave rise to what we called a topological chiral soliton \cite{Brown:1984sx}. We understood here the role of the negative energy quark sea restoring baryon number and energy  by  ``leaking"  through the bag surface. This phenomenon showed a remarkable insensitivity of these quantities to the bag radius, a signature, that could be extended to any observable. This interesting phenomenon  lead  Mannque to propose later  the so called ``Cheshire-Cat principle". My collaboration with Mannque had been so enjoyable and successful that after my thesis I decided to move to Saclay. There, under the charming atmosphere of Paris, I enjoyed not only the physics but also an intense personal relationship between our two families.

The next important stage of our professional relationship occurred when Byung-Yoon Park came as a postdoctoral fellow to Valencia. His presence strengthened the collaboration with Mannque and gave rise to the quantitative foundation of the Cheshire-Cat principle (CCP) \cite{Park:1989bk}. Years later, Hee-Jung Lee also became a postdoctoral fellow in Valencia, and together with Byung-Yoon and Mannque, we continued investigating the formalism and phenomenology of the chiral bag from the perspective of the CCP, addressing many problems among them the study of the proton spin a the light of the so-called "proton spin crisis" \cite{Lee:1998rj}.

In 1998, Mannque Rho obtained a professorship at KIAS (Korean Institute for Advanced Study), and the collaboration that had begun between Saclay and Valencia, moved to Seoul. Dong-Pil Min, professor at Seoul National University and thesis advisor of Byung-Yoon and Hee-Jung joined our investigations. In Seoul, the interests of the collaboration expanded to understanding nuclear matter and its properties through the study to Skyrme matter. The idea was to describe nuclear matter starting from the solitonic picture of baryons which is supposed to represent
QCD for large number of colors. For this purpose, the instanton-skyrmion connection of Atiyah and Manton was exploited to describe skyrmion matter \cite{Park:2002ie}.  
This research spanned more than ten years, thanks to Mannque's persistence and the excellent work of Byung-Yoon, with his deep understanding of the problem and his enormous ability to perform extremely complex calculations. The Skyrme model allowed us to study the transition to dense nuclear matter, something that was impossible for lattice QCD \cite{Park:2008zg}. From 2008 onward, Mannque pursued other directions, while still motivating from a distance my investigations with Byung-Yoon into matter at high densities and temperatures \cite{Park:2009bb,Park:2019bmi}. 

In the following sections, I will briefly describe some of the most fruitful ideas from the research I conducted with Mannque, initially at Stony Brook-Saclay, later at Valencia-Saclay, and finally at KIAS-SNU. The emphasis will be mostly on the initial ideas, not on the more complete later developments. I will not describe technical details nor aim for a complete bibliography. From the references mentioned, both the technical methods and a full set of citations can be obtained.

\section{The Little Bag} 
It is an attractive possibility that quark confinement can be described as a phase
transition phenomenon. Within this picture, the usual vacuum cannot tolerate
color fields. The quark structure of hadrons is therefore confined to certain regions
of space, called bags, where color fields are present. If
the bags are small enough, the quantum chromodynamical interaction of the quarks
becomes weak and can be treated perturbatively.  The original  MIT bag model~\cite{Chodos:1974je} had the 
disturbing feature that chiral symmetry was not preserved.  Implementing chiral symmetry
in the bag model was the leitmotiv that started  in 1978 our investigation regarding  
the structure of hadrons and nuclear matter.  In 
the chiral bag the quarks which are  confined generate an axial current at the surface which 
we  made continuos by introducing pion degrees of freedom outside the bag. The pions outside the bag create a pressure which make the 
confining bag size shrink, thus the popular name``little bag" \cite{Brown:1979ui}.
Let us be more specific. In the bag picture, quark wave functions do not vanish at the surface. Thus, the
quark contribution to the axial-vector current is discontinuous. Continuity is restored
by introducing pion degrees of freedom outside the bag in accordance with spontaneously
broken chiral symmetry.   A great advantage of working with the pion field is that it makes
possible to maintain the usual hypothesis of a partially conserved axial-vector
current (PCAC) as an operator identity, incorporating into the  model a
variety of successful soft-pion predictions. The fundamental new equation of the model at the bag surface consists in the coupling of  the pion fields $\Phi^i$ outside the bag
to the quarks $\Psi$ inside which is determined by the requirement of a conserved axial-vector current 

\begin{equation}
f_\pi (\vec{n} \cdot \vec{\partial} \, \Phi^i) = i \overline{\Psi} \gamma_5 \frac{1}{2}\tau^i \vec{n} \cdot\vec{\gamma} \Psi
\end{equation}
where $\vec{n}$ is an outward normal to the surface. 

The remaining equations for the chiral bag are direct chiral-symmetric 
generalizations of corresponding MIT bag equations \cite{Vento:1980mu}.

\begin{figure}[htb]
\begin{center}
\includegraphics[scale= 0.77]{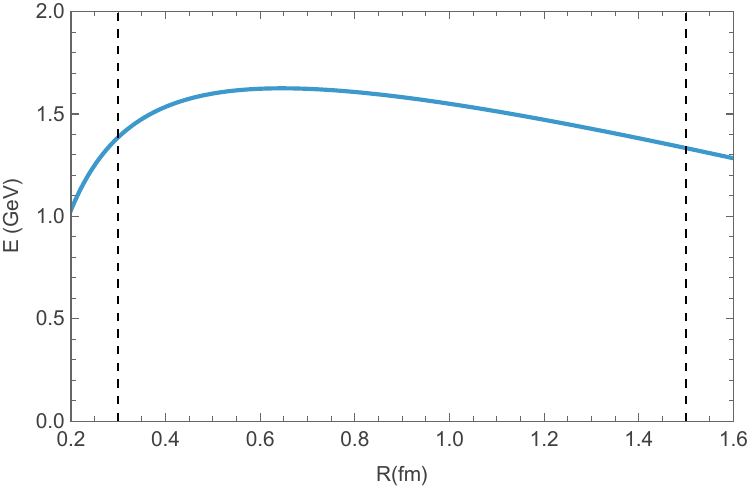} 
\end{center}
\caption{Energy of a nucleon-delta bag as a function of bag radius calculated in the spherical chiral bag model}
\label{EBag}
\end{figure}

\begin{figure}[htb]
\begin{center}
\includegraphics[scale= 0.77]{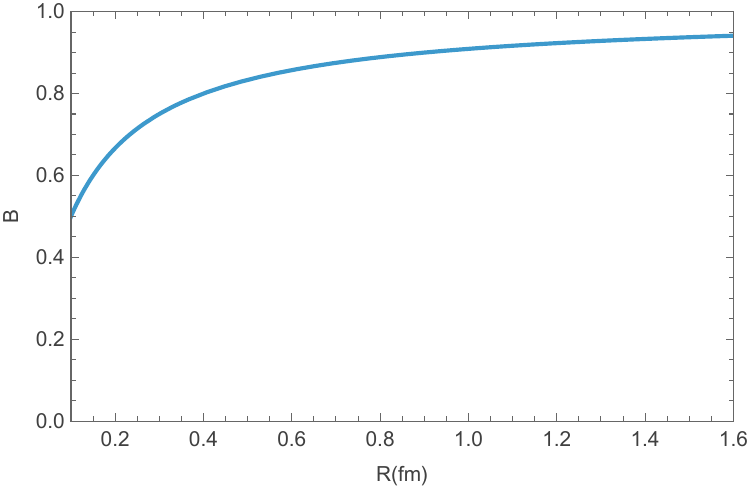} 
\end{center}
\caption{Baryon number of a $\Delta$-N bag as a function of bag radius calculated in the spherical chiral bag model}
\label{BBag}
\end{figure}

A spherical solution, the so called hedgehog ansatz, of the equations of the chiral bag model was found. There were many consequences of  this solution, which were discussed in those days  regarding chiral symmetry  \cite{Vento:1980mu}, but from todays perspective  the most important result was that most of the observables where quite independent of bag radius. In Fig. \ref{EBag} we show the bag energy as a function of radius signalling the region were the energy is almost independent of the bag radius. This lead to the so called Cheshire cat principle (CCP), which we will  discuss in the next section. An interesting result, which Mannque pointed out to me, arose from the calculation of the baryon number. As shown in Fig. \ref{BBag} the baryon number in our model becomes less the $1$ as the radius decreases. This controversy led us to understand  the solitonic nature of the pion and the role of the negative energy sea in the calculation of observables later on.

\section{The Cheshire Cat Principle}

Let us consider a version of the chiral bag model in which the interior quark sector is joined to 
an exterior solitonic meson sector through the requirement of continuity of the axial vector current at 
the bag surface. The negative  energy quark sea plays a crucial role in this model, which   
reduces to the Skyrme soliton~\cite{Skyrme61}  in the limit as the bag radius $R\rightarrow$ 0 \cite{Brown:1984sx}. 
The``leakage" of baryon number and energy through the bag results in a remarkable insensitivity 
of these quantities to the bag radius. This discovery lead to the development Cheshire cat principle (CCP)
in hadron structure. The CCP states that at low energies, physics should not depend
upon the confinement radius $R$. This is equivalent to saying that at low energies
QCD can be bosonized. If the CCP were to hold exactly, then all
the observables given in terms of quark variables could always be described in terms
of hadronic degrees of freedom. Since baryons arise as solitons in effective bosonic
lagrangian theories, this means that those observables can always be put in terms
 of bosonic fields. In (1 + 1) dimensions, this phenomenon can be rigorously established.
In fact, one can bosonize a certain class of fermion theories exactly and
show that physics is truly independent of the confinement radius. However in
(3+1) dimensions, this is not possible. In (3+1) dimensions there are no exact
bosonization rules and hence one can not establish a perfect CCP.
It can only be approximate. The goal of low-energy effective theories of
QCD is then to write a bosonized theory which maximally captures the essence of
QCD in the nonperturbative regime. The chiral bag provides a tool with which this task
can be carried out. 

\begin{figure}[htb]
\begin{center}
\vskip 0.5cm
\includegraphics[scale= 0.77]{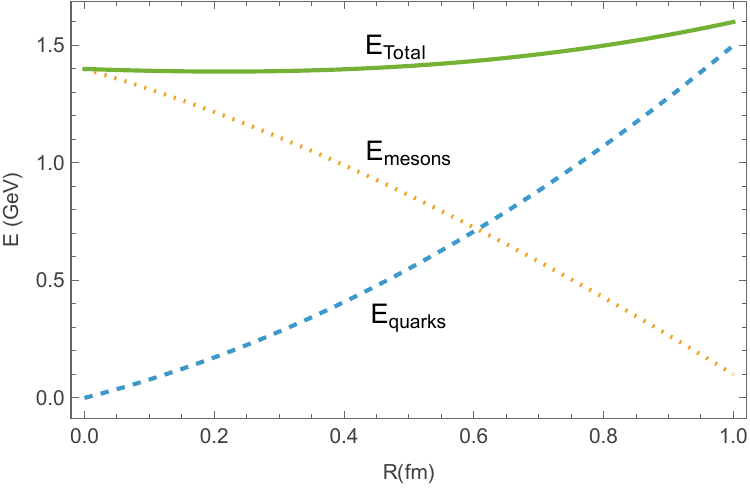} 
\end{center}
\caption{The energy of a baryon in GeV as a function of bag radius calculated in the chiral bag model. Individual contributions from quark 
and meson sectors are shown.}
\label{END}
\end{figure}

 For example one can calculate the energy of a baryon in the model as a function of $R$. 
 The result of this calculation is given  in Fig. \ref{END}, 
 which shows both the total energy of the baryon and the fractionation of energy between quark 
 and meson sectors.

For small $R$, the energy is dominantly in the mesons. In the limit $R\rightarrow 0$ all the energy is contained in the soliton. 
As $R$ increases, energy leaks through the bag into the quark sector. The sum of quark and meson energies remains remarkably 
constant over the range $0<  R < 1$ fm. These results show a tremendous similarity to the fractionation of baryon number \cite{Rho:1983bh}.

Another interesting example we studied was the proton matrix element of the flavor-singlet axial 
current (FSAC), often referred to as``proton spin",  in terms of the chiral bag model with due account 
of the axial anomaly and quantum number fractionization. We found  that independently of the bag
radius $R$, the FSAC matrix element remains small, say, less than $15\%$, as required by experiment. We suggested that  this is an
intricate manner in which the Cheshire cat principle manifests itself \cite{Park:1989bk}.

\begin{figure}[htb]
\begin{center}
\vskip 0.5cm
\includegraphics[scale= 0.77]{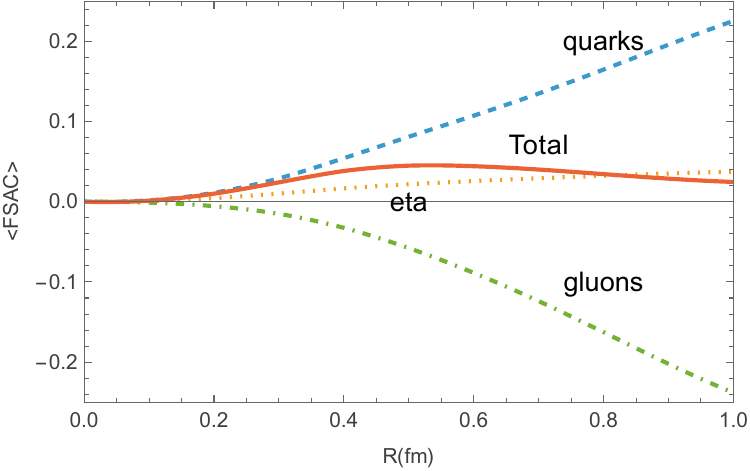} 
\end{center}
\caption{The proton expectation value of the FSAC in a chiral bag model as a function of bag radius. The value for $R=0$ corresponds to the result of the Skyrme model.}
\label{FSACag}
\end{figure}

A few years later  a more detailed account including Casimir effects involving gluons for the FSAC
as well as other phenomena (such as $\Delta-N$ splitting, flavor-singlet magnetic moment
of the nucleon, etc.) was reported in close agreement with experiment confirming  the CCP  \cite{Lee:1998rj}

Strictly speaking, the CCP in hadronic systems can not
persist indefinitely. It must break down eventually. Now the
major task in hadronic physics is to understand whether or not the signature of the
true QCD degrees of freedom could be explicitly seen in physical processes in the
phase we study. For this it is crucial to understand to what extent the Cheshire cat
phenomenon persists and at what circumstances it breaks down. The chiral bag provides
and ideal tool for that purpose.

\section{Skyrmion Matter}

The Skyrme model~\cite{Skyrme61} describes baryons as topological solitons 
of an effective meson Lagrangian, which can be interpreted 
as a large $N_c$ limit of QCD~\cite{Witten79}.
Once the topological winding number is identified with
the baryon number the model provides an excellent framework for hadron physics leading to a reasonable description of 
the properties of  baryons \cite{Zahed:1986qz}.

The baryon number one particles associated with this lagrangians are called skyrmions.
Due to its classical nature, which may be subject to quantization if required, 
the Skyrme model is also able to describe dense matter, the so called skyrmion matter,
free from any obstacles such as the ``sign problem" in lattice QCD. 
Moreover, the strong interaction generated by the skyrmions can be treated nonperturbatively. 
The great advantage of the model is that the same Skyrme Lagrangian 
can describe both infinite baryonic matter 
and its mesonic fluctuations  in a unified way~\cite{Lee:2003aq}.     

\begin{figure}[htb]
\centering
\vskip -1.5cm
\includegraphics[scale= 0.5]{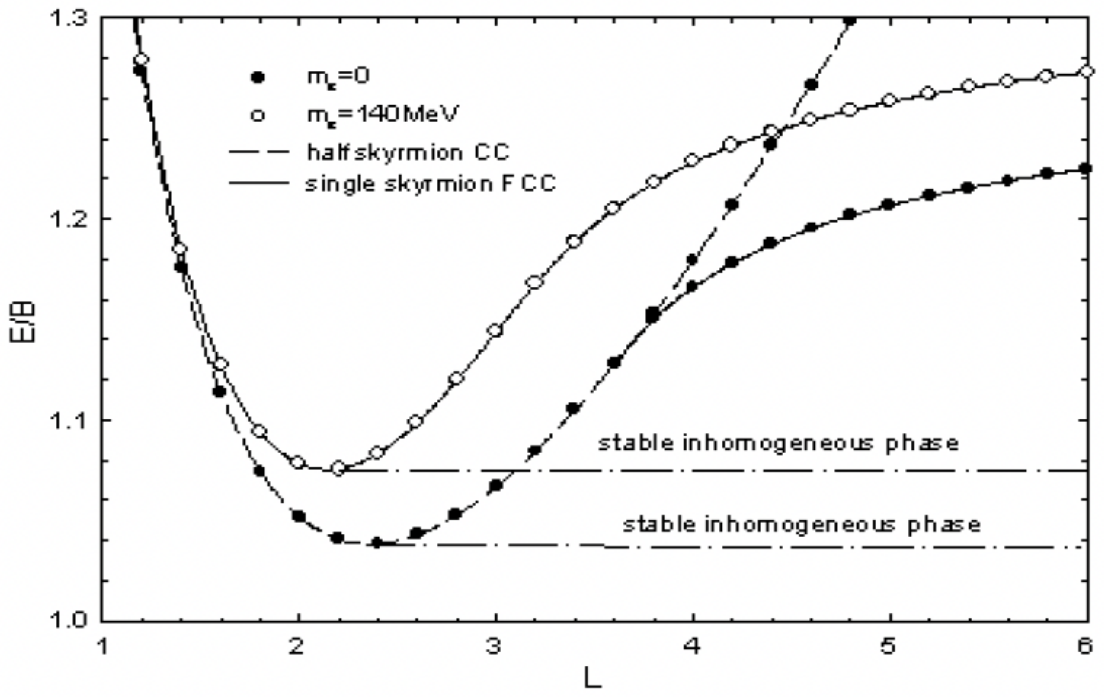} 
\vskip -2.5cm
\caption{Energy per single skyrmion as a function of the size parameter $L$. The solid circles
show the results for massless pions and the open circles are those for massive pions. Note the
rapid phase transition around $L \sim 3.8 $for massless pions}
\label{phasetransition0}
\end{figure}

Our  interest has been  in understanding the phase transitions of skyrmion matter and the associated changes in the hadron properties. Skyrmion matter 
undergoes a phase transition from the FCC (Face Centered Cubic)  single skyrmion configuration to the CC  (Cubic Crystal) half-skyrmion configuration.
This can be seen in  Fig. \ref{phasetransition0} where we show the calculation in our model of the energy per baryon $E/B$ as a function of the FCC box size parameter
$L$. Note that the baryon density is related to this parameter by $\rho= 1/2L^3$. Each point denotes the minimum energy of the crystal configuration for a given value
of $L$. The solid circles correspond to the zero pion mass calculation and reproduce
the results of Kugler and Shtrikman~\cite{Kugler:1989uc}. In order for an easier comparison we present the results
with $L$ given in units of $(ef_\pi)^{-1} $ ($\sim 0.45$ fm) and $E/B$ in units of $(6\pi^2f_\pi)/e$ ($\sim 1160 $ MeV),
respectively. The latter enable us to compare the numerical results on $E/B$ easily with its
Bogolmoln'y bound for the skyrmion in the chiral limit, which can be expressed as $E/B = 1$ in
this convention.  

In  Fig. \ref{phasetransition0} one sees that as we squeeze the system from $L = 6$ to around $L = 3.8$,
the skyrmion system undergoes the mentioned phase transition from the FCC single skyrmion configuration
to the CC half-skyrmion configuration. The system has a minimum energy at $L \sim 2.4 $ with
the energy per baryon $E/B \sim 1.038$, which is very close to the Bogolmoln'y bound.
Note in Fig. \ref{phasetransition0} that in the region $L > 3.8$, two different phases coexist. This phase transition occurs in the unstable region where
the system has negative pressure, therefore it does not have any physical consequence.
The phase to the left of the minimum is referred to as ``homogeneous" and there
the background fields are described by a crystal configuration. The phase to the right of the
minimum is ``inhomogeneous" since the pressure $P = \partial E/\partial V$ is negative, the skyrmion matter
is unstable, and to make it stable we will assume a realization in terms of an ``inhomogeneous"
phase, where the skyrmions condense to form dense lumps with large empty spaces.

\begin{figure}[htb]
\vskip -1.5cm
\hskip  0 cm
\includegraphics[scale= 0.6]{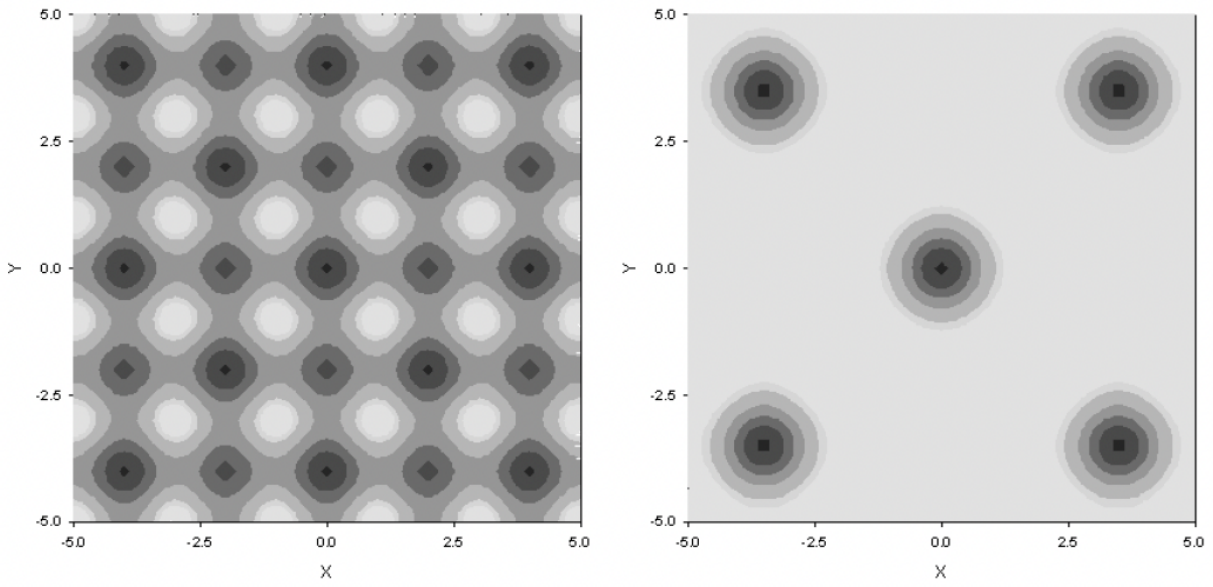} 
\vskip -4.5cm
\caption{Local baryon densities. The figure on the right is the FCC crystal made of single skyrmions. The figure of the left represents an CC lattice of localized almost half skyrmions.}
\label{phasetransition1}
\end{figure}

The open circles are the solutions found with a nonvanishing pion mass, $m_\pi = 140$  MeV.
Comparing to the skyrmion system for massless pions, the energy per baryon is somewhat higher.
Furthermore, there is no first order phase transition at low densities~\cite{Lee:2003aq}.

To see the configuration of the final skyrmion crystal, we present in Fig. \ref{phasetransition1} the baryon
density on the x-y plane (z = 0) at two different values of density. When it is dilute (right), the system is composed of well separated single
skyrmions lodged on each FCC lattice site. At high density (including the minimum-energy configuration), the individual
skyrmion loses its identity because of the overlap among the skyrmions. The dense centers
of the single skyrmion split into two and the final configuration forms a CC crystal whose
unit cells carry one-half unit of baryon number.

In all cases, when matter is made of half-skyrmions, the value of the half-skyrmions at the lattice sites is such that 
the average value of  the non linear field $U = \exp(-i \vec{\tau}\cdot \vec{\pi}/f_\pi)$, vanishes.   
The vanishing of $\langle U \rangle$  was interpreted as a symptom 
of  chiral symmetry restoration in a dense medium. 
However, in ref.~\cite{Lee:2003aq}, it was pointed out that  $\langle U\rangle =0$ does not mean 
the vanishing of the pion decay constant in medium and therefore does not imply the restoration of chiral symmetry. 
The mechanism that takes place is well known under the name of ``pseudo-gap" scenario.
At higher density $f_\pi$ does vanish and chiral symmetry is restored.
Thus the chiral restoration phase transition in the most naive Skyrme model happens in a two step process.
This unconventional realization is resolved by incorporating a dilaton field into the model Lagrangian, 
which is associated with the scale anomaly of QCD~\cite{Lee:2003aq}. 
Moreover, by incorporating more mesons such as the $\rho$ and the $\omega$ into the model Lagrangian via a hidden local gauge symmetry, the description improves resembling closely the standard scenario\cite{Park:2008zg} leading to quantitative predictions on physical quantities such as the critical density of the phase transition \cite{Ma:2013ela}.

However, a fundamental question on the phase transition remains.
From the very beginning the conventional approach adopted to study skyrmion matter 
assumes as ansatz for the minimum energy configuration a homogeneous crystal structure with certain symmetry. 
Although it is far from the Fermi liquid phase, which is generally accepted as a state for the normal nuclear matter, 
this solid crystaline structure is a reasonable starting point. 
However, the problem in this approach is that the phase transition occurs at a density 
where the pressure of the system is negative. 
Thus  the system is unstable against collapse and some mechanism should be provided to keep the system 
in that homogeneous phase. 
As suggested in ref.~\cite{Park:2002ie} at this density, where the system has negative pressure, the homogeneous 
symmetric configuration has no advantage in comparison to some  inhomogeneous phase where the skyrmions that are close to each other 
form lumps and leave large portions of space empty. This reasoning is supported by astrophysics where it is known that nuclear matter 
inside a neutron star appears in inhomogeneous phases 
known due to their structure by pasta names: gnocchi, spaghetti, lasagna, antispaghetti and
antignocchi~\cite{Caplan:2016uvu}. 
As a first step in the description of inhomogeneous skyrmion matter,
we constructed mathematically by adopting  the Atiyah-Manton ansatz~\cite{Atiyah:1989dq}  a few  planar structures made of skyrmions \cite{Park:2002ie}.  In the last paper of these series, a work in which Mannque did not participate directly, but liked and has cited repeatedly,  by adopting again the  Atiyah-Manton ansatz, we constructed a few layers of ``skyrmion sheets'' as a model for inhomogeneous skyrmion matter~\cite{Park:2019bmi}.

\section{Conclusions}

In this brief essay, I have traced the path I walked alongside Mannque Rho and many other mutual friends in physics. Chiral symmetry in all its manifestations has been the leitmotif of  this journey. From the early bag models to the more sophisticated soliton models, chiral symmetry has guided the steps of our understanding of low-energy physics, which we have tried to connect, through bag boundaries or large N approximations, to ultraviolet physics. I have presented three early manifestations of our work, where the basic ideas are more perceptible than in later, more sophisticated studies. These ideas have marked an important part of my scientific career, the part I spent collaborating with Mannque Rho. My account was meant to be a tribute to his ninetieth birthday. Unfortunately, Mannque left us before that, and this account has turned into a tribute to his memory, as an inspiring physicist and a friend. Throughout his career, Mannque sought a precise description of low-energy hadronic physics and its transition to quark and gluon physics. To that end, he used all the resources of non-perturbative physics, some of which this essay touches upon. The leitmotif in our research was always chiral symmetry, which is always associated with pions. In a certain way, the small bag revived the forgotten physics of Skyrme, as did large-N-color physics, and with that revival came intuitively exciting descriptions such as the CCP and nuclear physics based on solitonic meson models. 
Mannque was an inquisitive physicist always in search of new formalisms to represent the complexity of hadronic processes. He surrounded himself with young people whom he could excite with his ideas. He was also a great motivating force for Korean physicists. I fondly recall the celebration of his eightieth birthday in Pohang. Many of the talks by Korean physicists had him as their inspiration, and the atmosphere was one of admiration and respect. For me, that was the last time I had the chance to interact with him personally, and I glimpsed in him the same enthusiasm I had discovered back in 1978 when I first met him.
I want to conclude by expressing my enormous appreciation for Mannque as a physicist and as a person. To me, he was always a magnificent mentor at first, and later an excellent companion, along this long journey that is life. Thanks to him, I came to know an extraordinary country, Korea, where I spent cherished moments and made great friends.

\section*{Acknowledgements}
I would like to express my gratitude to the editors  for the opportunity given to me to honor Mannque Rho. I would also like to thank my close collaborators Hee-Jung Lee, Dong-Pil Min, and Byung-Yoon Park who were fundamental in carrying out the research reported here. Support by Grant CNS2022-135688 Consolidaci\'on Investigadora is acknowledged.

\end{document}